\documentclass[apj]{emulateapj} 
\usepackage{graphicx}  
\usepackage{dcolumn}   
\usepackage{bm}        
\usepackage{amssymb}   
\usepackage{color}
\usepackage{amsmath}
\usepackage{mathrsfs}

\def\apj{ApJ}
\def\apjl{ApJL}
\def\apjs{Astrophys. J. Supp. Ser. }

\def\aap{Astron. Astrophys. }

\def\mnras{MNRAS}

\def\nat{Natur}

\def\icarus{Icarus}

\def\be{\begin{equation}}
\def\ee{\end{equation}}
\def\ba{\begin{eqnarray}}
\def\ea{\end{eqnarray}}
\def\go{\mathrel{\raise.3ex\hbox{$>$}\mkern-14mu
             \lower0.6ex\hbox{$\sim$}}}
\def\lo{\mathrel{\raise.3ex\hbox{$<$}\mkern-14mu
             \lower0.6ex\hbox{$\sim$}}}

\begin{document}

\title{Shedding Light on the Eccentricity Valley: Gap Heating and Eccentricity Excitation of Giant Planets in Protoplanetary Disks}
\author{David Tsang}\email{dtsang@physics.mcgill.ca} \affiliation{Department of Physics, McGill University, Montreal, QC, H3A 2T8, Canada}
\author{Neal J. Turner}\affiliation{Jet Propulsion Laboratory, California Institute of Technology, Pasadena, CA 91109, USA}
\author{Andrew Cumming}\affiliation{Department of Physics, McGill University, Montreal, QC, H3A 2T8, Canada}

\date{\today}

\begin{abstract}
We show that the first order (non co-orbital) corotation torques are significantly modified by entropy gradients in a non-barotropic protoplanetary disk. Such non-barotropic torques can dramatically alter the balance that, for barotropic cases, results in the net eccentricity damping for giant gap-clearing planets embedded in the disk. We demonstrate that stellar illumination can heat the gap enough for the planet's orbital eccentricity to instead be excited. We also discuss the ``Eccentricity Valley'' noted in the known exoplanet population, where low-metallicity stars have a deficit of eccentric planets between $\sim 0.1$ and $\sim 1$ AU compared to metal-rich systems \citep{Dawson2013}. We show that this feature in the planet distribution may be due to the self-shadowing of the disk by a rim located at the dust sublimation radius $\sim 0.1$ AU, which is known to exist for several T Tauri systems. In the shadowed region between $\sim 0.1$ and $\sim 1$ AU lack of gap insolation allows disk interactions to damp eccentricity. Outside such shadowed regions stellar illumination can heat the planetary gaps and drive eccentricity growth for giant planets.  We suggest that the self-shadowing does not arise at higher metallicity due to the increased optical depth of the gas interior to the dust sublimation radius. 
\end{abstract}

\section{Introduction}

The rapid pace of extrasolar planet discovery in recent years has provided detailed statistical data about the distribution of planets in our Galaxy. One of the many surprises in these data is the large number of planets with significant eccentricity measured through the radial velocity (RV) method \citep{Butler2006}, particularly in those planets orbiting beyond $\sim 0.1$ AU, where tidal damping is inefficient. In contrast, the planets in our own Solar System (aside from Mercury) have low-eccentricity orbits. While eccentricity has been primarily measured through RV, eccentricity can also be inferred for the many exoplanets discovered through the transit method by utilizing transit timing variations (TTVs) \citep{Lithwick2012} and through the so-called `photoeccentric effect' on the transit lightcurve \citep{Dawson2012a}. \citet{Kane2012} recently showed through a comparison of orbital and transit timescales that the statistical eccentricity distribution of the planets detected by the \emph{Kepler} spacecraft may be similar to the distributions of those that have direct RV measurements of their eccentricity, however \citet{Plavchan2012} point out that this depends on the uncertain stellar radii given for the Kepler sample, particularly for smaller stars. \citet{Kane2012} also showed that the mean eccentricity of \emph{Kepler} planet candidates increases with increasing planet size plateauing above Neptune-sized planets ($R_p \go 4 R_\Earth$). 

The planets are thought to have migrated into their present orbits. Migration processes fall into two broad classes, planet-disk interactions \citep{GT80}, where planets migrate smoothly through the disk;  and few-body interactions, where planets are moved inwards through interaction with perturbing or scattering bodies \citep[see e.g.][]{Rasio1996, Lin1997, Naoz2011}. Multi-body dynamical processes have been studied intensely, in part to explain the presence of hot Jupiter systems that were found by early RV observations. 
However, nearly a third of the likely planet candidates found by the \emph{Kepler} observatory are thought to be in co-planar multi-planet systems \citep{Lissauer2012}, indicating that their evolution was likely dominated by disk interaction. The results of disk migration also provide the initial conditions of the planet locations and eccentricity distribution for the multi-body interactions that take place after the disk has dissipated. Given an initial distribution of sufficiently non-zero eccentricities ($e\go 0.1$), \citet{Juric2008} showed that dynamical relaxation of the planetary systems could explain the eccentricity distribution of higher eccentricity systems.

Migration driven by planet-disk interactions has been investigated as a means to excite eccentricity in migrating planets. Disk resonances exchange both energy and angular momentum between a planet and the disk, modifying the semi-major axis and the eccentricity of the planet. Certain resonances tend to pump eccentricity, while others tend to damp it. It is generally thought that the net effect of disk-planet interactions is to damp eccentricity. For small planets, this is almost certainly the case \citep{Artymowicz1993}. For giant planets, \citet{GS03} and \citet{OL03} suggested that saturation of the eccentricity-damping first-order corotation resonances could lead to eccentricity growth of planets able to clear a gap in the disk. However, recent numerical studies \citep{Dunhill2013, Bitsch2013b} have not seen such saturation in simulations, finding that eccentricity is damped for giant planets except for extreme cases where the gap is wide enough to avoid the corotation resonances entirely.

Entropy gradients in the disk can modify corotation resonances through baroclinic effects. Previous studies on the effects of entropy gradients in non-barotropic disks have focussed on the primary (co-orbital) co-rotation resonance, numerically solving for the response of the disk \citep{BM08, Paardekooper2008, Paardekooper2010, Paardekooper2011}, and the effect of the non-barotropic non-linear horseshoe torques on Type I migration of small planets embedded in the protoplanetary disk. These non-barotropic effects can allow halting or even reversal of migration for certain systems. \citet{Bitsch2013a} recently utilized the numerical fitting formulae from \citet{Paardekooper2011} to predict the effect of stellar illumination on migration of small embedded planets. 

In a companion paper \citep{Tsang2013a}, we go beyond these previous studies by deriving a fully analytic form for the linear non-barotropic corotation torques, for both primary and first-order corotation resonances. Here, we apply these results to eccentricity excitation in disks in which the gap formed by a migrating planet is heated by stellar irradiation. The general picture is as follows.  As a gap clearing planet perturbs the disk, Lindblad resonances excite the planet's eccentricity while non co-orbital corotation torques damp its eccentricity. In a barotropic disk, the damping dominates unless the gap is very large. We show here that the entropy gradients at the edge of an irradiated gap act to reduce the non-barotropic corotation torque, allowing the eccentricity-exciting Lindblad resonances to dominate. The requirement for the gap to be irradiated in order to enable eccentricity growth in the disk offers a potential explanation for the lack of eccentric planets at intermediate orbital periods around low metallicity stars found by \cite{Dawson2013}. In low metallicity disks, heating of the disk at the dust sublimation radius may lead to shadowing of the disk in the region $\sim 0.1$--$1\ {\rm AU}$ \citep{Dullemond2001}, causing eccentricity damping in this ``Eccentricity Valley'' region. 

An outline of the paper is as follows. In Section 2 we discuss the nature of non-barotropic corotation torques and their dependence on background entropy gradient.  In Section 3 we review the mechanism of eccentricity damping for giant planets in barotropic disks. In Section 4 we show that for reasonable assumptions about the disk structure and insolation the resulting non-barotropic torques can lead to eccentricity excitation. In Section 5, we demonstrate how the ``Eccentricity Valley'' may be explained by eccentricity damping and excitation in self-shadowed and fully irradiated disks respectively. Finally we discuss the implication of our results in Section 6.

\section{Corotation Torques in Non-Barotropic Disks}

\subsection{The Vortensity Equation}
A quantity that plays an important role in the determination of the corotation torque is the vortensity, $\zeta \equiv \omega_z/\Sigma$, (also known as the potential or specific vorticity) where $\omega_z = {\bm \hat{z}} \cdot( {\bm \nabla} \times {\bm u})$ is the z-component of a fluid's vorticity. \cite{GT79} showed that the torque associated with the corotation resonance has the sign of the radial vortensity gradient in the disk. For disks with orbital angular velocity $\Omega(r)$,  $\zeta$ can be written as
\be
\zeta = \frac{\kappa^2}{2\Omega \Sigma} = \frac{2B}{\Sigma},
\ee
where $\kappa^2 \equiv (2\Omega/r) d(r^2\Omega)/dr$ is the square of the radial epicyclic frequency, $B = \kappa^2/(4\Omega)$ is the local Oort constant, and $\Sigma$ is the disk surface density.

The continuity equation and the curl of the momentum equation for ideal fluids can be combined to obtain the Lagrangian (co-moving) time derivative for vortensity \citep[the vorticity equation, see e.g.][]{Lovelace1999},
\be
\frac{D}{Dt}\zeta = {\bm \hat{z}} \cdot \frac{{\bm \nabla} \Sigma \times {\bm \nabla} P}{\Sigma^3}. \label{vorticityeqn}
\ee
The right hand side of the equation above is the baroclinic term, which contributes when the lines of constant (surface) density $\Sigma$ and (vertically integrated) pressure $P$ are not parallel in a fluid.

In most previous works studying eccentricity evolution due to disk-planet interaction, the disk is assumed to be barotropic, such that pressure is only a function of density, $P = P(\Sigma)$. This causes the lines of constant $P$ and $\Sigma$ to be parallel for any perturbations, such that the baroclinic term is zero. Thus for barotropic disks, the vortensity is conserved for all perturbations. 

In a non-barotropic disk, where entropy gradients exist, the baroclinic term is non-zero. For an ideal gas with adiabatic index $\gamma$ where $S \equiv P/\Sigma^\gamma$, we can write Equation \eqref{vorticityeqn} as
\be
\frac{D}{Dt}\zeta = {\bm \hat{z}} \cdot \frac{{\bm \nabla} \Sigma \times {\bm \nabla} S}{\Sigma^{3-\gamma}}, \label{vorticity2}
\ee
which shows that density perturbations may generate vortensity in the presence of an entropy gradient.

\subsection{Disk Resonances}

A planet orbiting with non-zero eccentricity has a time varying gravitational potential which can be split into the Fourier components
\be
\Phi_p(r) = \sum_{l,m} \Phi_{l,m} \cos[m(\phi + \Omega_{l,m} t)],
\ee
where $m\Omega_{l,m} = m\Omega_p + (l-m)\kappa_p$ is the pattern frequency for the ($l,m$) component, and $\Omega_p$ and $\kappa_p$ are the orbital and epicyclic frequencies of the planet. Only the $l=m$ (primary) and $l = m\pm 1$ (first-order), components are non-zero to first order in eccentricity. 

At Lindblad resonances, where the pattern frequency of the gravitational perturbation in the frame co-moving with the disk material is equal to the local epicyclic frequency ($m(\Omega_{l,m} - \Omega) = \pm \kappa$), the disk material is able to respond strongly to such perturbations and waves are launched, resulting in Lindblad torques. 

At a corotation resonance, the disk material moves with the same speed as the pattern speed for a particular potential component. The primary corotation resonances (where $\Omega(r) = \Omega_{m,m} = \Omega_p$) are located at the same orbital radius as the planet, while the first-order corotation resonances (where $\Omega(r) = \Omega_{m\pm1,m} = \Omega_p \pm \kappa/m$) are located both interior and exterior to the orbit of the planet (see Figure \ref{ResonanceLocation2}). In the remainder of this paper we will use the terms ``first-order corotation resonance'' and ``non co-orbital corotation resonance'' interchangeably.

\subsection{The Corotation Torque}

A full calculation of the linear non-barotropic corotation torque must also consider the disk enthalpy response to the perturbing potential. This has been previously studied for co-orbital (primary) corotation resonances by numerically calculating this response \citep{BM08, Paardekooper2008, Paardekooper2010, Paardekooper2011}. In the companion paper to this work, \citet{Tsang2013a}, we employ an approach similar to that of \citet{ZL06} to derive a fully analytic form for the linear non-barotropic corotation torque for both primary and first-order corotation resonances. 

While the full expression from \citet{Tsang2013a} should be used to evaluate such torques in the general case, for the purposes of this discussion we will adopt the simplifying limit of a cold ideal gas disk, such that the adiabatic sound speed, $c_s^2 \equiv \gamma P/\Sigma \ll r^2 \Omega^2$, is small compared to the orbital velocity (such that  the disk scale height, $h/r \equiv c_s/r\Omega \ll 1$, is small). We allow the disk to have entropy gradients such that there can be a non-zero radial Brunt-V\"ais\"ala frequency, $N_r$, defined to be
\be
N_r^2 \equiv -\frac{1}{\gamma \Sigma} \frac{dP}{dr} \frac{d \ln S}{dr}.
\ee
For simplicity we also assume that $N_r^2 > 0$ such that the disk is stably stratified radially, though this may not be the case for some disks, particularly those where the baroclinic instability only occurs slowly \citep{Lesur2010}, or where magnetic fields have prevented instability from occurring \citep{Lyra2011}. For the gaps examined later in the paper $N_r^2 > 0$ near the gap edges, as both the pressure and entropy gradients change sign.

For $N_r^2/\Omega^2 \ll 1$ the expression for the corotation torque of the disk acting on the planet can be given as \citep{Tsang2013a}
\footnote{Here we have also taken the assumption that the background entropy and density gradients are roughly comparable and not too extreme, $|d\ln S/d\ln r| \sim |d\ln\Sigma/d\ln r| < (h/r)^{-1}$, which is not unreasonable for the models we consider later. For sharper entropy profiles or thicker disks the full expression of Equation 47 in \citet{Tsang2013a} should be used. In sharp gaps where $d\ln\Sigma/d \ln r \sim (h/r)^{-1}$ the 3-dimensional torques should be considered \citep[see e.g.][]{ZL06, Tanaka02}.}
\be
\Gamma^{\rm (CR, cold)}_{l,m} \simeq -\left[\frac{2 m \pi^2 \Sigma \Phi_{l,m}^2}{(d\ln\Omega/dr) \kappa^2} \right]_{r_c} \frac{d}{dr}\ln \zeta_{\rm eff} + {\cal O}(N_r^4/\Omega^4) \label{torqueeqn},
\ee
where the ``effective'' vortensity is given by 
\be
\zeta_{\rm eff} = \zeta/(S^{1/\gamma}) = \frac{\kappa^2}{2\Omega \Sigma S^{1/\gamma}},
\ee
and $r_c$ is the radial location of the corotation resonance. 

The sign of the corotation torque for a cold disk is thus controlled by the combination of the radial gradients of the vortensity and the disk entropy. In the limit where $N_r^2 \rightarrow 0$ and $S \rightarrow$ constant, we recover the original expression for barotropic the corotation torque from \citet{GT79} (see their eq.~[56]), where the sign of the torque depends only on the gradient of the vortensity. Note that if the characteristic timescale of energy transport in the disk is shorter than the perturbation (orbital) timescale, then the perturbations cannot be treated as adiabatic, but will be isothermal, and the barotropic torque will also be recovered.

\subsection{Understanding the Sign of the Corotation Torque}

\begin{figure}
\includegraphics[width=\columnwidth]{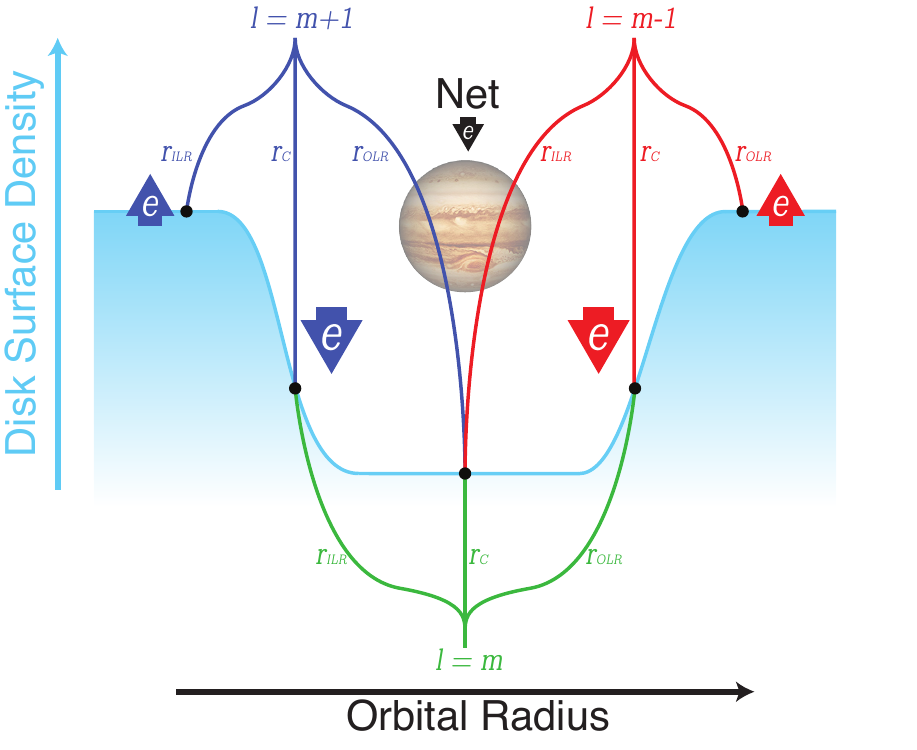}
\caption{Schematic illustration for eccentricity damping by disk-planet interactions in a gap for a barotropic (or isentropic) disk. The location of the corotation ($r_c$), inner Lindblad ($r_{\rm ILR}$) and outer Lindblad ($r_{\rm OLR}$) resonances are shown relative to the gap-clearing planet for the $\Phi_{l,m}$ primary ($l=m$) and first order ($l = m\pm 1$) potential components. Also shown are the dominant contributions to the eccentricity evolution assuming a cleared gap. The eccentricity damping by the non co-orbital corotation torques tends to slightly dominate (by $\sim 5\%$) the eccentricity pumping by the first-order external Lindblad resonances \citep{GS03} leading to a net eccentricity damping by the disk-planet interaction. \label{ResonanceLocation2}}
\end{figure}

The nature of the corotation torque can be understood qualitatively by considering the effect of the potential components on collisionless particles near the corotation point \citep{GS03}. Just interior to the corotation resonance a particle feels a more slowly moving potential which slows it down and moves it towards the corotation radius.  Just outside the corotation a particle feels a more quickly moving potential, which speeds it up and moves it towards corotation. There is therefore a net particle flux towards the corotation, with the particle starting interior to the corotation gaining angular momentum and those starting exterior to the corotation losing it. The net torque on disk (and conversely on the planet) then depends on the relative flux of material into corotation from either side. 

For barotropic perturbations the vortensity is conserved by Equation \eqref{vorticityeqn}, such that the vortensity is carried with fluid elements regardless of the torques exerted by the planet. In order to have the net vortensity of the incoming material match the vortensity at the corotation the relative flux towards the corotation from either side must scale with the background vortensity gradient $d\zeta/dr$, and thus the net torque at a corotation resonance must be proportional to this value for barotropic disks.

However, for disks where an entropy gradient exists, such density and pressure perturbations can generate vortensity. The flux of material towards the corotation from either side needed to match the net vortensity at the corotation must be altered by the baroclinic term of the vortensity equation, thus the net torque is modified to include the effect of the entropy gradient.

\begin{figure}
\includegraphics[width=\columnwidth]{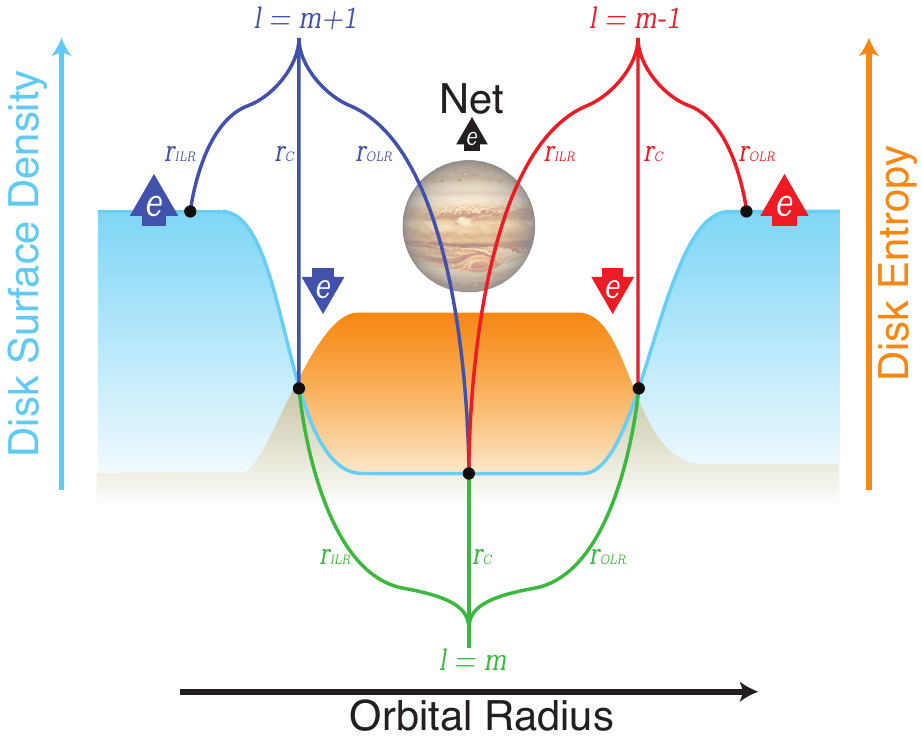}
\caption{Schematic illustration for eccentricity excitation by disk-planet interaction in a gap for a non-barotropic disk with a  heated gap. The location of the corotation ($r_c$), inner Lindblad ($r_{\rm ILR}$) and outer Lindblad ($r_{\rm OLR}$) resonances are shown relative to the gap-clearing planet for the $\Phi_{l,m}$ primary ($l=m$) and first order ($l = m\pm 1$) potential components. Also shown are the dominant contributions to the eccentricity evolution assuming a cleared gap. The presence of an entropy gradient at the non co-orbital (first order) corotation resonances can cause a reduction (or even a reversal) of the corotation torque compared to the barotropic case (see Figure \ref{ResonanceLocation2}), allowing the first-order external Lindblad resonances to dominate and thus a net eccentricity excitation by the disk-planet interaction. \label{ResonanceEntropy2}}
\end{figure}

\section{Eccentricity Damping in Barotropic Gaps}

\subsection{Eccentricity Evolution Through Disk Torques}
Planets exchange both energy and angular momentum with the disk through the resonances at different rates which allows eccentricity evolution of the orbit to occur \citep{GT80, Artymowicz1993, GS03}.

The eccentricity of a planet, $e_p$, can be given in terms of its energy $E_p$ and its angular momentum $H_p$,
\be
e_p^2 = 1 + \frac{2 H_p^2 E_p}{(M_p)^3 (GM_\star)^2}.
\ee

The transfer of energy and angular momentum between the planet and the disk at any resonance occurs due to interaction of the potential component $\Phi_{l,m}$, with the disk material. This potential component is constant in frame corotating with $\Omega_{l,m}$, so that the Jacobi constant $J_{l,m} = E_d - \Omega_{l,m} H_d$ of the disk material is conserved, where $E_d$ and $H_d$ are the energy and angular momentum of the disk. Consequently the resonance transfers energy and angular momentum between the disk and the planet such that
\be
\frac{dE_p}{dt} = \Omega_{l,m} \frac{dH_p}{dt} = \Omega_{l,m} \Gamma_{l,m},
\ee
where $\Gamma_{l,m}$ is the torque acting on the planet. 
Combining these equations to lowest order in eccentricity we find \citep{GS03, GT80}, 
\be
e_p \frac{de_p}{dt} = (\Omega_{l,m} - \Omega_p) \frac{H_p^2 \Gamma_{l,m}}{(GM_\star)^2 M_p^3} \label{ecceqn}.
\ee

External Lindblad torques (those not coincident with the planet location, see Fig 1) tend to drive eccentricity growth (and migration), while co-orbital Lindblad torques damp the eccentricity of the planet \citep{GT80}. The co-orbital Lindblad resonances tend to dominate for small planets by a factor of $\sim3$ \citep{Ward86, Artymowicz1993}, thus disk-interaction tends to damp eccentricity of small embedded planets.

\subsection{Planets In Barotropic Gaps}

If a planet is massive enough to clear a gap, such that the density of the co-orbital material falls dramatically, the damping by the co-orbital Lindblad resonances is reduced. \citet{GS03} then found that in a barotropic disk, for small eccentricity, the dominant terms\footnote{\citet{WardHahn2000} note that waves launched due to the apsidal resonance should significantly damp eccentricity even when a gap exists. \citet{GS03} argue that the long wavelength of the apsidal waves allows wave reflection to occur \citep[see e.g.][]{Tsang2011}, significantly reducing the amplitude of the apsidal torque when the waves are not resonant.} are the eccentricity excitation by the first-order external Lindblad resonances, and the eccentricity damping by the non co-orbital corotation torques (see Figure \ref{ResonanceLocation2}), due to the vortensity (density) gradients in the gap.

For first-order corotation resonances located at the gap edge \emph{interior} to the planet orbit, the positive vortensity gradient (negative density gradient) leads to a negative corotation torque on the planet ($\Gamma_{l,m} < 0$), in the barotropic limit of Equation \eqref{torqueeqn}. Since $\Omega_{l,m} > \Omega_p$ interior to the planet location, Equation \eqref{ecceqn} yields eccentricity damping.  

For first-order corotation resonances located at the gap edge \emph{exterior} to the planet orbit, the negative vortensity gradient (positive  density gradient) leads to a positive torque ($\Gamma_{l,m} > 0$). However, $\Omega_{l,m} < \Omega_p$ exterior to the planet's orbit, and eccentricity is again damped.

\citet{GS03} found that the full unsaturated corotation torques tended to dominate by a small amount ($\sim 5\%$) leading to a net eccentricity damping, but along with \citet{OL03} noted that only a small amount of saturation of the corotation was necessary to allow net eccentricity excitation to occur. 

Whether this condition for eccentricity growth is met depends in detail on the non-linear interactions that dictate the corotation saturation. Numerical simulations \citep[see e.g.][and references therein]{Dunhill2013} have shown that for some disk and planet parameters eccentricity growth can be seen. In particular for extremely large planets where a very large gap has been cleared, such that no material is present to interact at the non co-orbital corotation resonances, eccentricity growth driven by the first-order Lindblad resonances is seen \citep{Papaloizou2001, KD06}\footnote{In interpreting their simulation, \citet{KD06} refer to the ``damping effect of the outer 1:2 Lindblad resonance'', which we understand to mean the damping effect of the non co-orbital corotation resonance located at the same radius as the 1:2 Lindblad resonance.}.

However, \citet{Dunhill2013} recently performed a series of high-resolution 3-D smooth particle hydrodynamic simulations for various planet and disk parameters, and found that for realistic planet mass and disk properties planet eccentricity tends to be damped (aside from the systems with extremely large gaps discussed above), implying that insufficient saturation of the non co-orbital corotation resonances occurs in realistic systems to excite eccentricity. This led \citet{Dunhill2013} and others \citep[see e.g.][]{Dawson2013} to conclude that disk interactions cannot be a significant source of the eccentricity currently observed in many exoplanets. \citet{Bitsch2013b} recently performed several simulations calculating the damping rates for eccentricity and inclination. We note, however, that all of these simulations were performed using a locally isothermal equation of state and without considering the effect of  stellar illumination on the thermal structure of gap, i.e.~the gaps were effectively barotropic.

\section{Eccentricity Excitation in Heated Gaps}

\subsection{Gap Heating by Insolation}

Stellar illumination of the disk can significantly alter its geometrical structure and temperature profile \citep[see e.g.][]{Chiang1997}.  In particular, \citet{Varniere2006}, \citet{Turner2012} and \citet{Jang-Condell2012} have used Monte Carlo and semi-analytic radiative transfer codes to show how insolation modifies the thermal structure around a gap opened by a planet.  All three of these works treat radiative transfer in an axisymmetric model disk.  In the latter two, the hydrostatic vertical structure is also iterated to consistency with the radiation field.  If the disk has a flaring shape and the gap is optically-thin, the star directly illuminates and heats a  strip of material along the top of the gap's outer edge.  The outer  edge in turn emits infrared photons, some of which travel down into  the gap, lifting temperatures there above the equilibrium levels  found with no gap.  The gap's temperature is nearly uniform and determined by the balance between the received infrared photons and  the radiative cooling, proportional to the Planck-mean opacity of the material in the gap.  

\begin{figure}
\includegraphics[width=\columnwidth]{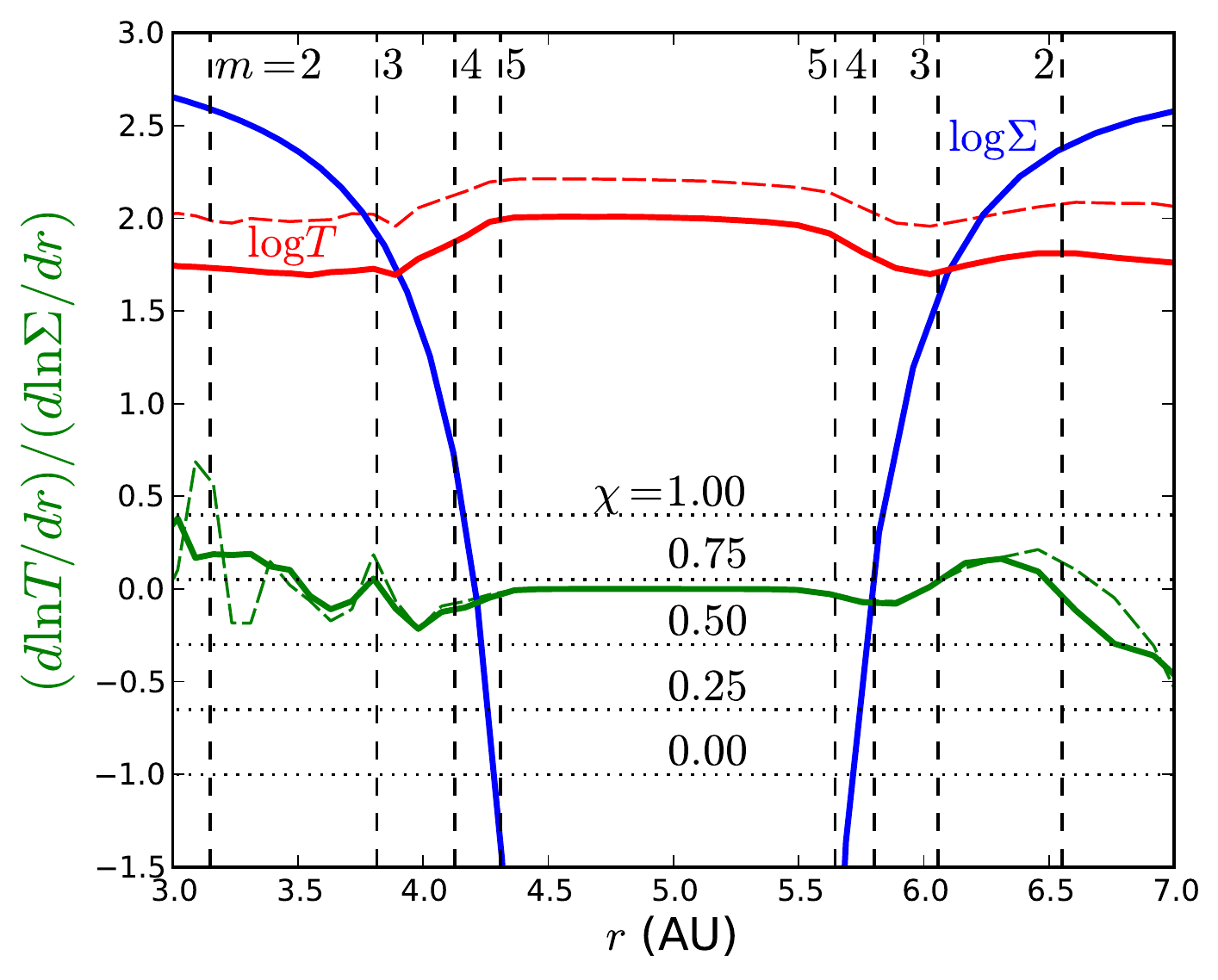}
\caption{The disk's surface density $\Sigma$ (blue; in units of g cm$^-2$) and temperature $T$ (red; in units of K) profiles for a gap-clearing $1M_J$ planet located 5 AU from a $1M_\odot$ star with luminosity $1L_\odot$ (solid) or $10L_\odot$ (dashed) from the Monte Carlo radiative transfer calculations of \citet{Turner2012}. The green curves show the ratio of the characteristic density and temperature lengthscales, $(d\ln T/dr)/(d\ln\Sigma/dr)$. Dotted lines are also shown corresponding to torque ratio $\chi \equiv \Gamma/\Gamma_{S\rightarrow {\rm const}}$ of various values, assuming adiabatic index $\gamma = 1.4$, for a diatomic ideal gas. We see that while the entropy gradients at the gap edges are insufficient to reverse the sign of the corotation torques ($\chi$ remains greater than zero), they are enough to significantly reduce the torques, with $\chi \lo 0.75$ in the region of the gap edges. The vertical dashed lines show the locations of the $m=2,3,4,5$ non co-orbital corotation resonances, assuming a Keplerian rotation profile in the disk. The fluctuations in the logarithmic derivative ratio at and inside the interior m=2 resonance seemingly imply $\chi > 1$ (and thus $N_r^2 < 0$),  however, these fluctuations are due to Monte Carlo noise from the low packet counts in the previous steps of the model iteration, and likely to shrink if more packets were used. 
\label{TSigmaplot}}
\end{figure}

The temperature profile of a gap depends on the geometry of the disk, in particular the angle of incident stellar illumination, and the amount of luminosity intercepted by the outer gap edge. As a specific example, Figure \ref{TSigmaplot} shows the surface density and midplane temperature profiles for two models with a Jupiter-mass planet 5 AU from a Solar-mass star.  The parameters are identical to the viscous model disk and 4600~K star from \citet{Turner2012}, but we repeated the calculations increasing the number of photon packets ten-fold to $10^8$ on the fifth iteration, to ensure that the temperature gradients at the gap edges are adequately sampled. The gap structure was obtained using the prescription from \citet{Lubow2006}, with viscous stress parameter $\alpha=0.005$, planetary accretion efficiency $E=6$, gap width parameter $f=2$ and planet/star mass ratio $q=1/1000$. The luminosities in the two models span the range expected between the beginning ($10L_\odot$) and end ($1L_\odot$) of the planet-forming epoch.

\subsection{Corotation Torque in a Heated Gap}
In a non-barotropic disk, the presence of an entropy gradient at the gap edge can modify the balance of eccentricity excitation and damping (see Figure \ref{ResonanceEntropy2}). In Equation \eqref{torqueeqn} the corotation torques are modified by the presence of an entropy gradient with the torque proportional to the logarithmic derivative of the effective vortensity, 
\be
\frac{d\ln \zeta_{\rm eff}}{dr} = \frac{d}{dr}\ln\left(\frac{\kappa^2}{2\Omega}\right) - \frac{d\ln\Sigma}{dr} - \frac{1}{\gamma}\frac{d\ln S}{dr}.
\ee
The torque due to an isentropic ($S\rightarrow {\rm const}$) disk background is identical to the torque for a barotropic disk. 
For an ideal gas disk with adiabatic index $\gamma$ and temperature $T$,  we can write the logarithmic entropy gradient of the background as
\be
\frac{d\ln S}{dr} = \frac{d\ln T}{dr} - (\gamma - 1) \frac{d\ln\Sigma}{dr},
\ee
giving
\be
\frac{d\ln \zeta_{\rm eff}}{dr}= -\frac{3}{2r} - \frac{d\ln\Sigma}{dr}\left[\frac{1}{\gamma} + \frac{1}{\gamma}\frac{d\ln T/dr}{d\ln \Sigma/dr} \right].
\ee
For a gap edge we can expect $|d\ln\Sigma/d\ln r| \gg 3/2$,  which allows us to estimate the ratio of the non-barotropic and barotropic torques as,
\be\label{eq:chi}
\chi \equiv \frac{\Gamma^{\rm (CR)}_{l,m}}{[\Gamma^{\rm (CR)}_{l,m}]_{S\rightarrow{\rm const}}} \simeq \frac{1}{\gamma} + \frac{1}{\gamma}\frac{d\ln T/dr}{d\ln \Sigma/dr}.
\ee
In an isentropic disk, $S$ is constant such that the temperature and density gradients must be related by
\be
\left(\frac{d\ln T/dr}{d\ln\Sigma/dr} \right)_{S \rightarrow {\rm const}} = \gamma - 1,
\ee
and $\chi\rightarrow 1$, and the torque is the same as the barotropic value. However as we now show, in the case of a heated gap $\chi$ can be significantly lower than 1.

\begin{figure*}
\includegraphics[width=\textwidth]{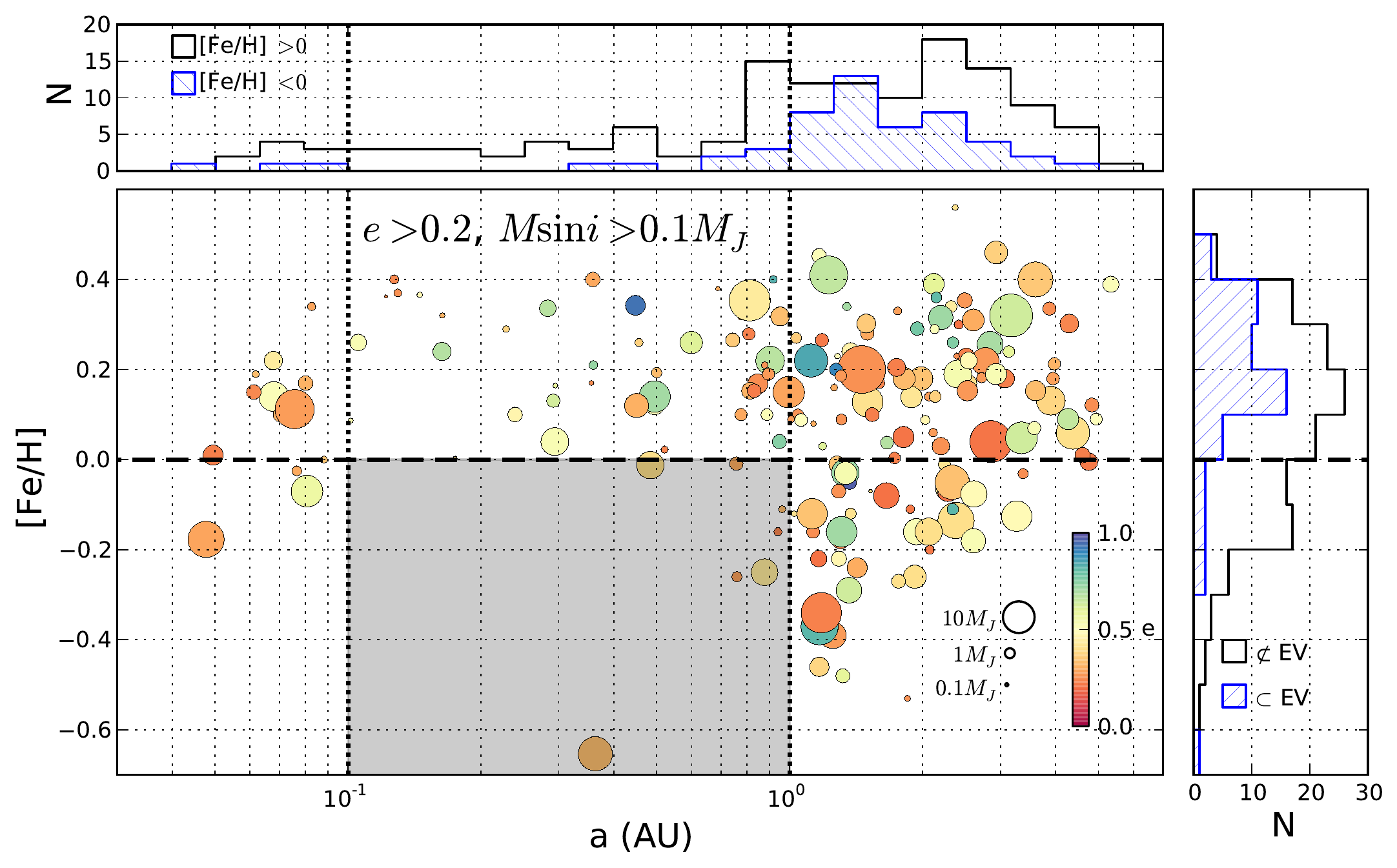}
\caption{Data from the Exoplanet Orbit Database \citep{Wright2011} illustrating the ``Eccentricity Valley'' for planets around low-metallicity stars \citep{Dawson2013}. Here we have plotted the stellar metallicity, [Fe/H], and semi-major axes, $a$,  of all RV measured planets with $M\sin i > 0.1 M_J$ and eccentricity $e > 0.2$ that have metallicity measurements listed. The size of each circle is proportional to $M\sin i $, while the color corresponds to the eccentricity with red being low and blue being high eccentricity. The ``Eccentricity Valley'' is the shaded region between 0.1 and 1 AU, where \citet{Dawson2013} found a strong deficit of eccentric planets around metal-poor host stars. For the upper plot  the hatched blue histogram represents the number of low metallicity ([Fe/H] $< 0$) planets in a particular radius bin while the other histogram is the number of high metallicity planets ([Fe/H] $> 0$). For the right plot, the hatched blue histogram represents the number of planets inside the ``Eccentricity Valley'' (EV) region ($0.1$ AU$ < r < 1$ AU) in a particular metallicity bin, while the solid histogram represents the number of planets outside this region.  \citet{Dawson2013} were able to reject with 99.14\% confidence that the lack of high eccentricity planets in the ``Eccentricity Valley'' for metal poor stars is by chance.
\label{EccData1}}
\end{figure*}

\citet{GS03} estimated that in a barotropic disk the eccentricity damping by the non co-orbital corotation torques exceeds the eccentricity pumping by first-order Lindblad torques by $\sim 5\%$. The Lindblad resonances are not significantly affected by the presence of entropy gradients as their locations are only perturbed slightly so that the perturbation frequency in the co-moving frame, $m(\Omega_{l,m} - \Omega)= \pm \sqrt{\kappa^2 + N_r^2}$, matches the epicyclic frequency modified to include the effect of the radial buoyancy.  If we take $\chi_{\rm crit} \sim 0.95$ to be the corotation torque reduction factor necessary to allow eccentricity driving by the Lindblad resonances to win, then we arrive at the condition for eccentricity driving by the disk, 
\be
\frac{d\ln T/dr}{d\ln\Sigma/dr} < \gamma\, \chi_{\rm crit} - 1. \label{critheat}
\ee
When the radial temperature scale height is much larger than the radial density scale height ($\left| d\ln T/dr/d\ln\Sigma/dr\right|\ll 1$), as is the case in the almost isothermal heated gap, we expect a reduction in the torque by a factor of $\chi\approx 1/\gamma$. An even smaller torque can result when the temperature gradient is of the opposite sign to the density gradient (if there is a strong temperature inversion going into the gap).

This result assumes that there is sufficient thermal diffusion to prevent the torque from thermally saturating. This is discussed further in \citet{Tsang2013a}, where it is shown that the dimensionless diffusivity required at the location of the first-order corotation resonances is small. The minimum dimensionless diffusivity required to maintain linearity is 
\be
D_e \equiv {K\over \rho C_p r_c^2 \Omega^2} > 10^{-7} \left(\frac{e}{10^{-2}} \frac{M_p/M_\star}{10^{-3}}\right)^{3/2},
\ee
for the model gap width and location used in \citet{Turner2012}, where $K$ is the thermal conductivity, $C_p$ is the specific heat at constant pressure, $\rho$ is the volume density, and all quantities are evaluated at the resonance location. If the diffusivity is too small, then the gradient of the entropy perturbation becomes comparable to the background entropy gradient, and the torque thermally saturates, returning to its barotropic value.  The minimum thermal diffusivity $D_e$ is exceeded by several orders of magnitude near the gap we consider here. 

To show a specific example, we have plotted the logarithmic derivative ratio for the gap models discussed in \S 4.1 in Figure \ref{TSigmaplot}, where we see that the insolation into the gap heated it enough to provide $\chi \lo 0.75$ at the gap edge. In fact, the heating of the gap has reduced the $(d\ln T/dr)/(d\ln \Sigma/dr)$ term in equation (\ref{eq:chi}) so much that the value of $\chi$ is close to $\chi\approx 1/\gamma=0.71$.

This level of gap heating can be achieved for systems where the gap width and disk flaring angle allow the stellar luminosity absorbed by the outer edge of the gap to be at least comparable to the heating due to accretion. \emph{Thus we expect that many giant planets experience eccentricity excitation by the disk  when they are relatively close to the star and have their gaps exposed to insolation}, in contrast to the results of previous barotropic simulations \citep[e.g.][]{Dunhill2013}. If the light from the central star is blocked, or the angle of incident starlight is sufficiently small (in, for example, a disk where the dust has significantly settled), then the entropy gradients will be much less steep, and the corotation torque should be closer to the barotropic level for which the planets will experience eccentricity damping due to the disk. 

The gaps involved must be sufficiently clean to suppress the damping of eccentricity by the co-orbital Lindblad resonances. Defining $\mathscr{C}_{CR}$, $\mathscr{C}_{eLR}$ and $\mathscr{C}_{cLR}$ to be the relative strengths of the non co-orbital corotation, external Lindblad, and co-orbital Lindblad resonances respectively \citep[c.f.][]{GS03}, we have $\mathscr{C}_{eLR}/\mathscr{C}_{cLR} \simeq 1/3$ \citep{Artymowicz1993}. Assuming a $\sim 5\%$ net eccentricity driving from the external Lindblad and corotation resonances, such that $\mathscr{C}_{eLR} - \mathscr{C}_{CR} \simeq 0.05 \mathscr{C}_{eLR}$, material in the gap will lead to eccentricity damping unless the gap is sufficiently clear $\Sigma_{\rm gap}/\Sigma_{\rm disk} \lo (\mathscr{C}_{eLR} - \mathscr{C}_{CR})/\mathscr{C}_{cLR} \simeq 1/60$.  

It should be noted that the final value of a planet's eccentricity cannot be determined through the linear analyses described above if the planet experiences eccentricity driving through the disk. Non-linear effects quickly become important as the eccentricity grows large and the planet nears the gap edge. However we note that even in barotropic simulations, once very wide gaps made the first-order corotation torques ineffective, the eccentricity
quickly reached large values \citep{Papaloizou2001, KD06, Dunhill2013}. Also, we assumed the disk is more massive than the planet.  Movement of material within the disk could limit the planet's final eccentricity in cases where disk and planet masses are comparable. 

\section{Disk Shadowing and the Eccentricity Valley}
\subsection{The Eccentricity Valley}
\citet{Dawson2013} recently showed that there is a puzzling deficit of eccentric planets with semi-major axis between 0.1 and 1 AU around metal-poor host stars (see Figure \ref{EccData1} for more details). By assuming that eccentricity can only be a result of a strong dynamical interaction from a third body, and noting the well known relationship between giant planet occurrence and metallicity \citep[see e.g.][]{Johnson2010}, they conclude that this metallicity dependent ``Eccentricity Valley'' is evidence that only systems with many giant planets can produce eccentric planets between 0.1 and 1 AU. 

However, as we have shown above, while a giant planet is embedded in the protoplanetary disk its eccentricity can be driven by interaction with the disk, provided its gap is sufficiently heated by the stellar illumination. This provides a means of exciting eccentricities even during the disk phase of the planet's orbital evolution, particularly for planets close to their host stars, as is the case for the bulk of the RV sample, and could provide the source of eccentricity for planets around metal rich stars. Below we suggest a possible mechanism for preventing this from occurring in metal poor stars located in the ``Eccentricity Valley'' region, through disk self-shadowing.

\subsection{The Dusty Inner Rims of T Tauri Disks, and Self-Shadowing}
\begin{figure}
\includegraphics[width=\columnwidth]{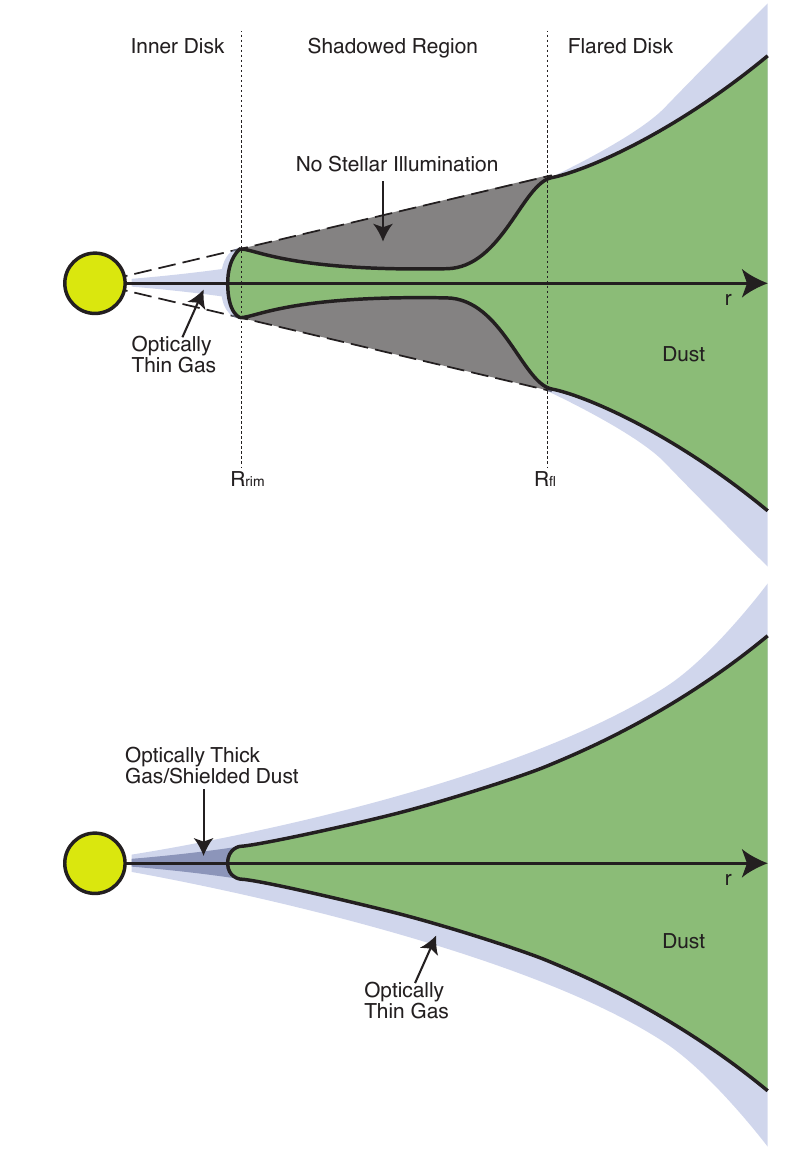}
\caption{\emph{Above}: A schematic diagram illustrating shadowing by a dust rim. If the gas interior to the dust sublimation radius is optically thin, then the disk at the sublimation radius ($R_{\rm rim}$) becomes puffed up as it receives insolation directly from the star, heating it far more than would be the case in a normal flared disk. This raised rim causes a shadowed region to form, where there is no direct stellar illumination of the disk. A giant planet located in this region does not have its gap heated by the central star. \citet{Muzerolle2003} showed that for T Tauri disks, the dust rim radius should be $R_{\rm rim} \sim 0.1$ AU.  The shadowed region should extend to the radius where insolation causes the disk to flare, $R_{\rm fl} \simeq 10-20 R_{\rm rim} \sim 1$AU \citep{Dullemond2001}. \emph{Below:} When the metallicity (or accretion rate)  is sufficiently high, the gas interior to the dust sublimation region is optically thick, and the disk has a `normal' flared shape \citep{Chiang1997}, which has gap heating at all radii.\label{DiskSchematic}}
\end{figure}

The progenitors of the solar mass stars studied by the RV sample in Figure \ref{EccData1}, are thought to be the classical T Tauri stars, which possess accretion disks from which planets will likely emerge. \citet{Muzerolle2003} showed that in a significant fraction of such stars a near-infrared excess exists, which is fit well by a single blackbody at the dust sublimation temperature. This was found to be consistent with a dust rim \citep{Dullemond2001, Natta2001} located at the dust sublimation radius. Fitting this model, \citet{Muzerolle2003} found that if the gas interior to the dust sublimation radius is optically thin \citep{Muzerolle2004}, then the disk at the sublimation radius, located at $R_{\rm rim} \simeq 0.1\ {\rm AU}\ (L_\star/L_\odot)^{1/2}(T_d/1500\ {\rm K})^{-2}$, where $T_d$ is the dust sublimation temperature, becomes puffed up as it receives insolation directly from the star, heating it far more than would be the case in a normal flared disk. This raised optically thick rim blocks the starlight, and causes a shadowed region to form, where there is no direct stellar illumination of the disk, as shown schematically in Figure \ref{DiskSchematic}. The disk is again illuminated by the star at $R_{\rm fl} \sim10-20 R_{\rm rim}$ \citep{Dullemond2001} where it begins to flare, yielding a range for the shadowed region of $\sim0.1 -  1$ AU, where embedded planets would have their eccentricity damped. 

The gas interior to the dust sublimation radius has opacity dominated by molecular lines (e.g. H$_{2}$O) between $\sim1500$K - $3000$ K \citep{Alexander1994, Eisner2009}. Thus for higher metallicities (or accretion rates) the gas interior to the dust sublimation radius can become optically thick, preventing the direct insolation and puffing-up of the dust rim. The disk should then have a normal flared shape \citep{Chiang1997}, which ``sees'' the star at all radii and has no shadowed region. A sufficiently flared disk would drive eccentricity growth for giant gap-forming planets embedded in the ``Eccentricity Valley'' region. 

\section{Discussion}

We have shown that entropy gradients and a non-barotropic equation of state modify the non co-orbital corotation torques for planets embedded in their natal disks. We have provided a simplified form for the torque valid in the limit of a cold disk, with $N_r^2 \ll \Omega^2$. A more general analytic formula for the corotation torque is provided in a companion paper \citep{Tsang2013a}. If starlight heats the gap around a giant planet enough to satisfy Equation \eqref{critheat}, then the corotation torques are reduced to the point where the planet's eccentricity is excited rather than damped.  This threshold is reached if enough of the stellar luminosity is absorbed on the gap's outer rim.  
We have shown that the threshold is easily exceeded in the flared
  structure obtained for a standard viscous disk model by placing the
  material surrounding the gap in radiative balance with the
  starlight.  Our starlight transfer calculations are axisymmetric and
  thus do not include the planet's wakes, but do capture the gross
  shape of the disk surface under vertical hydrostatic balance.  Thus
  we expect that starlight heating could have enabled many planets to
  gain eccentricity through disk interaction, rather than solely
  through later post-disk dynamical interactions as previously
  thought.  Note that eccentricity driving requires the gap around the
  giant planet to be sufficiently cleared that the co-orbital Lindblad
  torques are suppressed, since these would dominate the eccentricity
  evolution for an embedded planet  \citep[see e.g.][for further discussion]{GS03}.

The eccentricity driving has important implications for the inferred dynamical histories of giant planets in the currently known RV sample, and for the relative importance of orbital evolution due to the disk versus strong dynamical interactions, such as planet-planet scattering or the Kozai mechanism. Such disk interactions may be the source of the isolated eccentric giant planets.  For the \emph{Kepler} sample, the inferred mean eccentricity dependence on the planet size \citep{Kane2012}, rising with $R_p$ and plateauing above $4 R_\Earth$, may be consistent with an eccentricity growth process being tied to gap clearing conditions.  The eccentricity distribution at the end of the protoplanetary disk phase also provides the initial conditions for multi-planet interactions as discussed by \cite{Juric2008}, and further may help to explain the excess of planets with low, but non-zero eccentricity that are not explained by these interactions.

We have suggested that the ``Eccentricity Valley''  for planets between $\sim 0.1$ and $\sim 1$ AU around low-metallicity host stars \citep{Dawson2013} may be explained by noting that this range also corresponds to the region shadowed by the dust rim that is formed when an optically thin gas interior to the dust sublimation point allows direct insolation of the dust wall at the sublimation radius. In the self-shadowed region between $\sim 0.1$ and $1$ AU the disk interactions should damp eccentricity, resulting in circular orbits.  For higher metallicities, and higher accretion rates, the flow interior to the dust sublimation radius can become optically thick, preventing direct insolation from puffing up the dust rim, and causing the shadowed region to disappear. Such fully-illuminated disks may produce eccentric planets in the ``Eccentricity Valley'' region, as is seen for systems with [Fe/H] $> 0$. One might expect the extent of the shadowed region to decrease smoothly with increasing metallicity, as the height of the disk rim varies. Intriguingly, a hint of this may be present in the lower envelope of the planet distribution shown in Figure \ref{EccData1}.

We have only studied the linear growth of eccentricity in a system with a heated gap. To fully understand the eccentricity evolution of planetary systems and follow the eccentricity growth into the non-linear regime, detailed numerical simulations are needed that include non-barotropic equations of state and the effect of stellar illumination on the gap. Additional modeling of the dust rim and the shadowed region of protoplanetary disks, and the geometric effects of a planetary gap in these regions is also needed to confirm if the ``Eccentricity Valley'' arises due to self-shadowing and non-barotropic torques in fully-illuminated disks. 

Note that accretion heating is neglected in the two radiative transfer and hydrostatic balance calculations shown in Figure \ref{TSigmaplot}, as is shock heating associated with the planet's wake. Starlight will modify the corotation torques only when the gap is heated by illumination at least as much as the gap edges are heated by accretion and shocks. This would depend on the accretion rate at the end of the disk phase, and the gap opening dynamics and resulting structure.  It will also be important to study the expected entropy gradients in disks that are not irradiated or in regions of irradiated disks in which accretion heating dominates. 

\section*{Acknowledgments}
DT was supported by funding from the Lorne Trottier Chair in Astrophysics and Cosmology, and the Canadian Institute for Advanced Research. DT would like to thank Kostas Gourgouliatos, Philip Muirhead, Jason Wright, John Johnson, and particularly Rebekah Dawson for helpful advice and useful discussions during the course of this work. NJT carried out his part of the research at the Jet Propulsion Laboratory, California Institute of Technology, under a contract with the National Aeronautics and Space Administration and with the support of the NASA Origins of Solar Systems program under grant 11-OSS11-0074. AC was supported by an NSERC Discovery Grant and is an Associate Member of the CIFAR Cosmology and Gravity program.

\renewcommand{\bibsection}{\section{References}} 

\end{document}